\documentclass[
    ,final            % use final for the camera ready runs
  ]
  {aipproc}

\layoutstyle{6x9}

\usepackage{wrapfig}
\usepackage{floatflt}
\usepackage{caption}
\renewcommand{\figurename}{\small \bf FIGURE}

\newcommand{\pwo}{PbWO_4}

\newcommand{\sqrtsnn}{\sqrt{s_{_{NN}}}}
\newcommand{\pizero}{\ensuremath{\pi^0}}
\newcommand{\polpp}{\ensuremath{p^{\uparrow}p}}
\newcommand{\mgg}{\ensuremath{m_{\gamma\gamma}}}

\begin{document}

\title{Single Spin Transverse Asymmetries of Neutral Pions at Forward Rapidities
in $\sqrt{s}$~=~62.4 GeV Polarized Proton Collisions in PHENIX}


\author{Mickey Chiu for the PHENIX Collaboration}{
 address={Dept. of Physics, University of Illinois, Urbana, IL 61801 USA\footnote{\uppercase{P}resent address: \uppercase{B}rookhaven
 \uppercase{N}ational \uppercase{L}aboratory, \uppercase{U}pton, \uppercase{NY} 11375}}
}

\begin{abstract}
In the RHIC run of 2006, PHENIX recorded
data from about 20 $nb^{-1}$ of transversely polarized p+p collisions at
$\sqrt{s}$ = 62.4 GeV and polarization of about 50\%.  Also in this last
run PHENIX successfully commissioned a new $\pwo$ based
electromagnetic calorimeter, the Muon Piston Calorimeter (MPC),
covering $2\pi$ in azimuth and $3.1<\eta<3.7$.  The forward coverage
allows PHENIX to measure high $x_F$ $\pizero$  production.  
We present the current status of the analysis for the
single inclusive $\pizero$ transverse asymmetry $A_N$ at forward rapidities in PHENIX.
\end{abstract}

\maketitle

\section{Introduction}

Transverse single spin asymmetries in hadronic collisions have
had a long history of surprises, such as the observation by
the E704 collaboration of very large asymmetries in inclusive
pion production at high $x_F$ in $\sqrt{s}$ = 19.4 GeV transversely polarized
$\polpp$ collisions\cite{Adams:1991rw,Adams:1991cs}. Naively in leading twist perturbative QCD
one expects these asymmetries to be power suppressed such that $A_N \sim \alpha_{S}m_{q}/p_T$~\cite{Kane:1978nd}.
It was thought that at higher energies these asymmetries would be strongly suppressed.
Quite surprisingly, however, large asymmetries were discovered to persist even at collider energies by first STAR and
then the Brahms collaboration\cite{Adams:2003fx,Videbaek:2005fm}.

These asymmetries are interesting because they point toward some new understanding of internal
proton structure, such as the existence of a new non-perturbative function like the Sivers function\cite{Sivers:1989cc}, or
perhaps a non-zero transversity distribution in conjunction with a transversely dependent
fragmentation function\cite{Collins:1992kk} (Collins effect).  Of particular interest is that these effects
occur in a regime where NLO pQCD calculations correctly predict the unpolarized cross-sections.
Therefore the hope is that one can understand these asymmetries unambiguously, such as with a higher
twist expansion approach\cite{Qiu:1998ia,Koike:2002ti}, or by taking into account the transverse motion
of partons in a LO collinearly factorized approach with non-perturbative, transverse-momentum dependent functions\cite{Anselmino:2005sh}.  
Further progress in theoretical understanding requires more differential measurements such as the $p_T$ and $x_F$ dependence of the asymmetry, as well as 
di-hadron correlations\cite{Boer:2003tx} to distinguish between the above approaches.

\section{MPC Data Analysis and Current Status}

Before the 2006 run, PHENIX initiated a program to install a calorimeter inside the
hole at the front of the muon magnet piston\footnote{This idea was first suggested
by K. Imai and Y. Goto in 1996, and rediscovered recently. See http://www.phenix.bnl.gov/WWW/publish/goto/Polarimetry/pol1.html}
to extend the kinematic reach of PHENIX 
%\begin{floatingfigure}[rt]{0.6\textwidth}
\begin{wrapfigure}{r}{0.55\textwidth}
%\begin{figure}[ht]
 \centering
 \includegraphics*[width=0.5\textwidth]{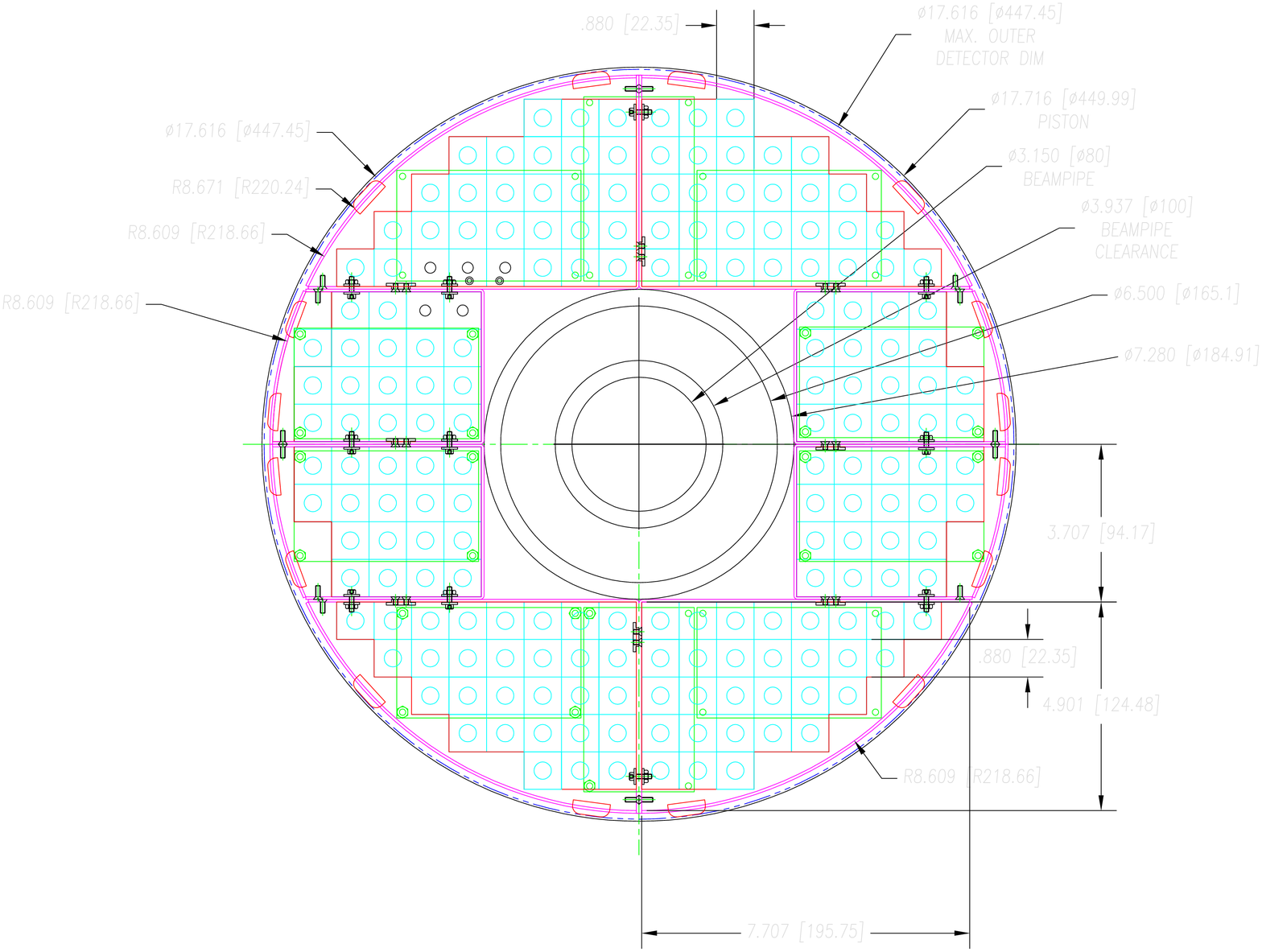}
% \noindent \hrulefill
 \renewcommand{\figurename}{\small \bf FIGURE}
 \caption{\small The PHENIX South MPC Layout.
 \label{fig:s_mpc}}
%\end{figure}
%\end{floatingfigure}
\end{wrapfigure}
calorimetry to more forward rapidities.
The muon piston holes are cylindrical with a depth of 43.1 cm and a diameter of 42 cm.  The small size of the area,
proximity to the interaction point, and sizable magnetic fields 
enforce tight constraints on the calorimeter's design, requiring a compact material with short radiation length
and small moliere radius and a readout that is insensitive to magnetic fields.  
PHENIX chose a design, based on that of the ALICE PHOS detector\cite{PHOS}, of a highly segmented
lead-tungstate crystal array with Avalanche Photodiode (APD) readout. 
Lead-tungstate is one of the best candidate materials for a compact calorimeter since it has one of
the smallest radiation length (0.89 cm) and moliere radius (2.0 cm) of any known scintillator.  PHENIX installed
192 crystals of size 2.2$\times$2.2$\times$18~$cm^3$ in the south piston hole in time for
the 2006 RHIC run.  The calorimeter sits around the beam-pipe 223 cm from the interaction point and covers $3.1<\eta<3.7$.  
The layout of the south MPC is shown in figure \ref{fig:s_mpc}.  Another 220 crystals will be installed in the north piston
hole for 2007.  Further details on the MPC hardware can be found in the contribution of A. Kazantsev
in these proceedings\cite{Kazantsev}.

\begin{floatingfigure}[r]{0.52\textwidth}
%\begin{wrapfigure}{l}{0.55\textwidth}
%\begin{figure}[ht]
 \begin{center}
 \includegraphics[width=0.48\textwidth]{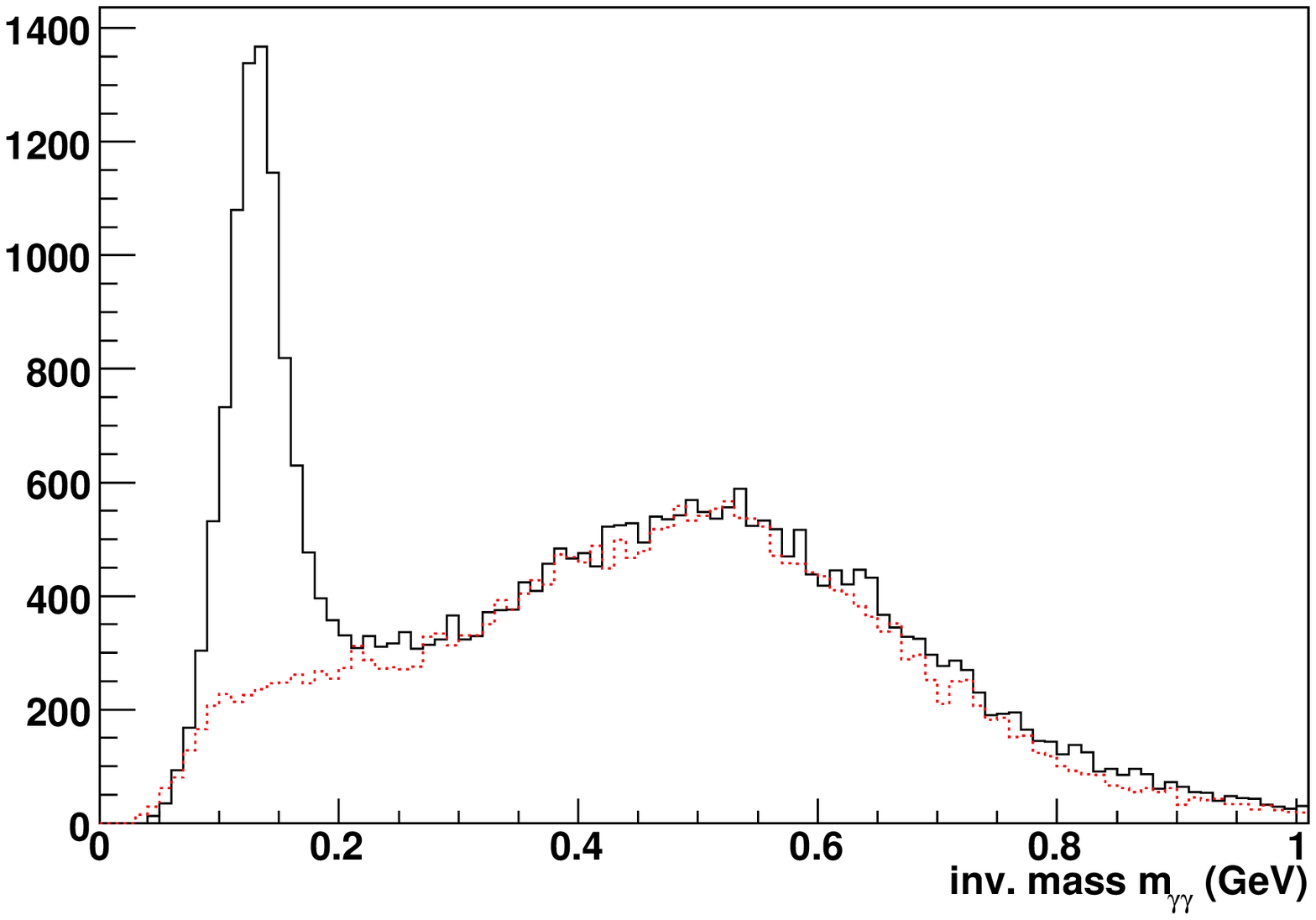}
 \end{center}
 \renewcommand{\figurename}{\small \bf FIGURE}
 \caption{\small The invariant mass distribution of photon pairs in the South MPC
 for p+p collisions at $\sqrtsnn=62.4$ GeV. The mixed event distribution is shown in
 the red dashed-dotted line.}
 \label{fig:inv_mass}
%\end{figure}
%\end{wrapfigure}
\end{floatingfigure}

The data shown here were taken with the PHENIX detector at RHIC during
two days of the 2006 $\polpp$ run, at a beam energy of $\sqrt{s}=62.4$ GeV.
A total integrated luminosity of $\sim$~20 $nb^{-1}$ of data were collected.
The data are triggered with the MPC using 4x4 tower energy sums at a threshold
of $\sim$~5 GeV, corresponding to good trigger efficiency for $\pizero$ at $\sim$
10 GeV.  An additional 80 $nb^{-1}$ of data were taken with longitudinally
polarized beams at the same beam energy\cite{Kazantsev}.

The MPC clustering algorithm is based on that developed for the PHENIX central arm calorimeter.
In this algorithm the known shower shape from test-beam measurements is used to 
reconstruct the hit location and energy of photons when the two photons are separated enough to see two peaks.
Alternative algorithms are being investigated which would have greater effectiveness at higher
energies, where the probability for the two decay photons to merge within the same
or adjacent towers is large, and it becomes difficult to separate one versus two photon showers.
The algorithms are evaluated by simulation.
In figure \ref{fig:inv_mass} the invariant mass distribution $\mgg$ of cluster pairs is shown,
along with pairs generated for clusters from mixed events.  Mixing clusters from different events
provides an estimate of the combinatorial background.  
Note that a region in the lower right quadrant of the MPC was excluded in this plot due to
issues with electronics noise.  Currently, the energy scale has been checked with the
MIP peak and $\pizero$ peak and both have been found to be consistent to
better than 10\%.

The single transverse spin raw asymmetry as a function of $\phi$ was calculated using the square-root
formula\cite{Ohlsen:1973wf,spinka}
\begin{equation}
{\cal A}(\phi) = \frac{1}{P_B}\epsilon_{N}(\phi) =
 \frac{1}{P_B}\frac{d\sigma^{\uparrow,\phi}-d\sigma^{\downarrow,\phi}}{d\sigma^{\uparrow,\phi}+d\sigma^{\downarrow,\phi}}
       = \frac{1}{P_B}\frac{\sqrt{N_L^{\uparrow,\phi}N_R^{\downarrow,\phi}}-\sqrt{N_L^{\downarrow,\phi}N_R^{\uparrow,\phi}}}
              {\sqrt{N_L^{\uparrow,\phi}N_R^{\downarrow,\phi}}+\sqrt{N_L^{\downarrow,\phi}N_R^{\uparrow,\phi}}}
\label{eqn:an}
\end{equation}
which largely cancels out differences in detector and beam asymmetries.
$\phi$ is the azimuthal angle relative to the beam polarization direction,
and $d\sigma^{\uparrow,\phi}$ represents the cross-section into the $\phi$ direction for a vertically
polarized beam with polarization $P_B$.  The asymmetry $A_N$ is then the amplitude of this asymmetry
modulation, ${\cal A}(\phi)=A_{N}sin(\phi)$.

In figure \ref{fig:raw_an} we have plotted on the left (right) the raw asymmetry $\epsilon_{N}(\phi)$
for photon pairs within the $\pizero$ mass peak for the cases where the yellow (blue) beam is
polarized.  In PHENIX the yellow beam is the proton beam heading towards the MPC, and the blue beam
is the one heading away.

\begin{ltxfigure}[ht]
%\begin{floatingfigure}[l]{0.98\textwidth}
%\epsfxsize=4.1in\epsfbox{c_anfit_correct.eps}
\vspace{3mm}
\centering
\includegraphics[width=4.4in]{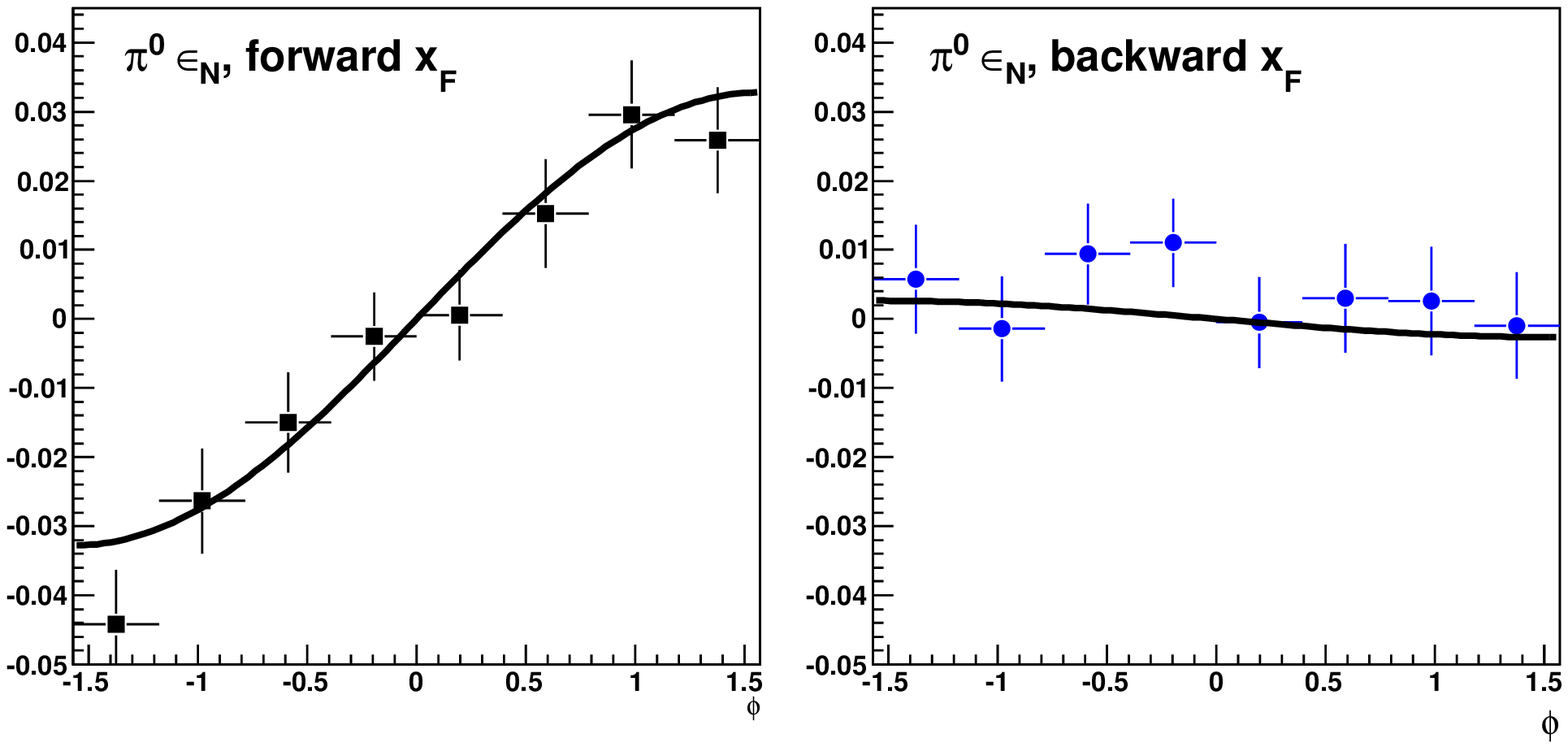}
\renewcommand\AIPfigurecaptionheadformat[1] {#1:\hspace{1em}}
\renewcommand{\figurename}{FIGURE}
\caption{\small Raw asymmetries from reconstructed photon pairs within $0.05<m_{\gamma\gamma}<0.25$~GeV
and $9<E_{\gamma\gamma}<12$~GeV.
The asymmetry has not been corrected for combinatoric background
or beam polarization.
\label{fig:raw_an}}
\end{ltxfigure}
%\end{floatingfigure}

There is a non-zero asymmetry for the case where the yellow beam is polarized, corresponding
to positive $x_F$ with respect to the polarized beam.  This asymmetry has been seen to increase
from low $x_F$ to higher $x_F$, and is well described by a $sin(\phi)$ functional form.  
For the case of the polarized blue beam (corresponding to negative $x_F$) the asymmetry is consistent with zero.
As an additional cross-check, the asymmetries
were found to be consistent with zero for the longitudinally polarized runs, as expected since then the
residual transverse polarization is small.

\section{Conclusions and Future Plans}

PHENIX has successfully commissioned a new forward electromagnetic calorimeter, the south MPC,
during the 2006 RHIC run, and plans to install the north MPC in time for 2007.  
With an early version of the clustering code, $\pizero$'s have been
identified in the MPC and a non-zero transverse asymmetry has been seen 
in $\polpp$ collisions at $\sqrt{s}$ = 62.4 GeV.  
Much work remains in improving the reconstruction code and in understanding any other artifacts of the data,
such as noise issues, backgrounds, and improving the determination of the energy scale,
before PHENIX cna finalize the cross-section and asymmetry results from the MPC.  
The validity of a perturbative theoretical interpretation of this data-set is currently under study.  
It is known that NLO pQCD fails to correctly describe the inclusive cross-section
correctly at these energies in the forward rapidities\cite{Bourrely:2003bw}.  However, including threshold resummation
should improve the agreement\cite{deFlorian:2005yj}.

Besides the transversely polarized data at 62.4 GeV, PHENIX also
collected 80 $nb^{-1}$ of longitudinally polarized data at $\sqrt{s}$ = 62.4 GeV
and another 7.5 $pb^{-1}$ of longitudinally polarized data
at $\sqrt{s}$ = 200 GeV during the 2006 run.  
The installation of the MPC allows PHENIX to confirm and contribute to previous measurements at high $x_F$ in RHIC collisions. 
Beyond mapping out the $x_F$ and $p_T$ dependence of $A_N$,
the high rate capabilities of the PHENIX DAQ will allow excellent triggering capability for
di-hadron correlation measurements between forward $\pizero$ and mid-rapidity hadrons.  Di-hadron measurements
are particularly interesting since they should help to distinguish between an asymmetry from an
initial state effect (Sivers) or a final state effect (Collins plus transversity).  Additionally,
di-hadron correlations constrain the kinematics of the initial scattered partons and therefore 
provide greater sensitivity to the sampled momentum fraction x of those partons.

\bibliographystyle{aipproc}
\bibliography{chiu_spin2006}

\IfFileExists{\jobname.bbl}{}
 {\typeout{}
  \typeout{******************************************}
  \typeout{** Please run "bibtex \jobname" to optain}
  \typeout{** the bibliography and then re-run LaTeX}
  \typeout{** twice to fix the references!}
  \typeout{******************************************}
  \typeout{}
 }

\end{document}